\documentstyle[12pt]{article}
\begin{document}

\title{Schr{\"o}dinger's Interpolation Problem through Feynman-Kac
Kernels }

\author{Piotr Garbaczewski \thanks{8th Symposium on Statistical Physics,
Zakopane, Sept. 25-30, 1995}\\
Institute of Theoretical Physics, University of Wroc{\l}aw,\\
PL-50 204 Wroc{\l}aw, Poland}

\maketitle
\hspace*{1cm}

\begin{abstract}
We discuss  the so-called Schr{\"o}dinger problem of deducing
the microscopic (basically stochastic) evolution that is consistent
with  given positive boundary probability densities for
a process covering a finite fixed time interval.
The sought for dynamics may preserve the
probability measure or induce its evolution, and is known to be
uniquely
reproducible, if the Markov property is required. Feynman-Kac type
kernels are the principal ingredients of the solution and determine
the transition probability density of the corresponding stochastic
process.
The result applies to a large variety of nonequilibrium
statistical physics and quantum  situations.
\end{abstract}

\section{Feynman-Kac kernels and time adjoint pairs of para\-bolic
equations  in  the description of  random dynamics}

The Schr\"{o}dinger problem \cite{schr} of reconstructing the
"most likely"  interpolating dynamics which is compatible with
the prescribed input-output statistics data
(analyzed in terms of nowhere vanishing  boundary probability
densities) for a  process with the time of duration  $T>0$,
can be given a unique solution, \cite{jam}.
For this purpose, it is necessary to define a suitable transition
probability $ m(A,B)=\int_Adx \int_Bdy m(x,y)$, mapping
among Borel sets $A\rightarrow B$ in time $T$, so that:
(a) the  bi-variate density $m(x,y)$ has the boundary data
$\rho _0(A), \: \rho _T(B)$ as its marginals for all $A$ and $B$,
(b) $ m(x,y)$ has the product form with a certain
strictly positive and jointly continuous in all variables kernel
as a  factor, \cite{jam,zambr}.
If the respective kernel is associated with a
strongly continuous dynamical semigroup, then the interpolating
process is Markovian, \cite{zambr,carm}.

The major issue, always  to be addressed  is:
 to specify under what circumstances (possibly
phenomenological, like in case of the boundary density data)
the kernel can be selected as appropriate, in reference to a
concrete physical situation.
Clearly, the obvious
and natural candidates with a direct physical appeal are the
familiar Feynman-Kac kernels, \cite{zambr,carm,blanch,olk}.

In the physical literature  a standard arena for the usage of
Feynman-Kac 
kernels, and the related Feynman-Kac representation formula for
solutions of  parabolic partial differential equations,  is either the
Euclidean  quantum theory \cite{glimm,simon,zambr,zambr1},
or the statistical physics of  nonequilibrium phenomena. For example,
the Fokker-Planck equation, with its non-Hermitian Markov generator,
is casually mapped into the parabolic evolution problem, whose
(semigroup)  generator is selfadjoint.
Indirectly \cite{carm,nag,blanch,olk,olk1}, through the
Cameron-Martin formula, the Feynman-Kac kernels  appear as an
important tool of the so  called stochastic analysis of measures 
and   related stochastic processes.  It is not 
accidental, since probability measures and their densities are 
involved in each of the considered frameworks  and studying dynamics 
in terms of densities \cite{lasota,karp} is a theory with much broader,
both physical and mathematical range of  applications,
than  indicated above.
 
The Schr\"{o}dinger equation and the generalized heat equation, which 
is basic for the original \cite{kac} Kac formula derivation, are
connected by analytic continuation in time. (We shall proceed in the 
notation appropriate to problems in space dimension one, although the 
main body of our arguments is space dimension independent).
For $V=V(x), x\in R$, bounded from below, the generator $H=-2mD^2
\triangle + V$ is essentially selfadjoint on a dense subset of 
$L^2$,  and the quantum unitary dynamics 
$exp(-iHt/2mD)$ is a final result of the analytic continuation
procedure  for  the holomorphic \cite{cast} semigroup
$exp(-H\sigma /2mD), \sigma =s+it, s\geq 0, s\rightarrow 0$.
Here, by equating $D=\hbar /2m$, the traditional notation
is restored.
Since the unit ball in $L^2$ is left invariant by 
the unitary  dynamics, and the Born \it statistical interpretation \rm 
postulate
assigns to  each normalized function $\psi (x,t)=[exp(-iHt/2mD) \psi ]
(x,0)$
a probability measure $\mu (A)=\int_A \rho (x,t) dx \leq 1$ 
with the density $\rho (x,t)= \psi (x,t)
{\overline {\psi}}(x,t)$, we are quite naturaly facing the problem of
the  existence of the random dynamics (stochastic process) which is
compatible  with the given time evolution of $\rho (x,t)$, or
preserves the measure in  the stationary case.

Let us emphasize that it is \it not \rm our goal to propose  any 
probabilistic 
"derivation" \cite{nel} of quantum theory.
Rather, we take seriously the 
Born postulate and attempt to draw  consequences of this assumption. 
Its impact is not  merely conceptual: the mathematical structure of
the theory is affected by
submitting the quantum unitary dynamics to the methods of stochastic 
analysis, appropriate for any standard probabilistic problem. 
Quite irrespectively  of whether we deal with essentially classical or 
quantum phenomena, and whether they are intrinsically random or 
have merely a random appearence (like in case of deterministic
derivations  of the stochastic, Brownian or Ornstein-Uhlenbeck type
evolutions, \cite{beck,grigo,karp}).

It is clear that the Madelung decomposition
$\psi (x,t)=[exp(R+iS)](x,t)$
of a \it nonzero \rm (on its domain of definition) solution of the 
Schr\"{o}dinger equation (we maintain the notation $D$ instead of 
$\hbar /2m$, and consider the conservative case $V=V(x)$):
$${i\partial _t\psi (x,t) = -D\triangle \psi (x,t) + 
{1\over {2mD}}V(x)\psi (x,t)}\eqno (1)$$
$$i\partial {\overline {\psi }}(x,t)=D\triangle {\overline {\psi }}
(x,t) - 
{1\over {2mD}}V(x){\overline {\psi }}(x,t)$$
where ${\overline {\psi }}=exp(R-iS)$ is a complex conjugate of
$\psi $, 
while $R(x,t), S(x,t)$ are real functions  and $\psi (x,0)$ is
taken  as the initial Cauchy data for equations (1),
implies the validity of the coupled pair of nonlinear partial 
differential equations
$$\partial _t \rho (x,t) = - \nabla (v\rho )(x,t)$$
$${Q(x,t) - V(x)= 2mD[\partial _tS + D(\nabla S)^2](x,t)}
\eqno (2)$$
where
$$\rho (x,t) ={\overline {\psi }}(x,t)\psi (x,t)=[exp(2R)](x,t)$$ 
$$v(x,t)=2D\nabla S(x,t)$$
$${Q(x,t)=2mD^2{{\triangle \rho ^{1/2}}\over {\rho ^{1/2}}}(x,t)}
\eqno (3)$$
If, instead of complex functions $\psi , {\overline {\psi }}$ we
introduce real functions (we follow  the notation
 of previous publications \cite{zambr,blanch,olk})
$$\theta (x,t)=[exp(R+S)](x,t)$$
$${\theta _*(x,t) =[exp(R-S)](x,t)}\eqno (4)$$
then equations (2) can be replaced by the (nonlinearly coupled via
$Q(x,t)$) pair of time adjoint parabolic
equations for $\theta (x,t), \theta _*(x,t)$:
$$\partial _t\theta _* =D\triangle \theta _* - 
{1\over {2mD}}(2Q-V)\theta _*$$
$${\partial _t\theta =-D\triangle \theta  +
{1\over {2mD}}(2Q-V)\theta }
\eqno (5)$$
with  the Cauchy data $\theta (x,0), \theta _*(x,0)$ fixed by the
previous Madelung exponents $R(x,0), S(x,0)$.
In turn, they imply  the validity of the equations (2) with 
$\rho (x,t)=[exp(2R)](x,t)=\theta (x,t)\theta _*(x,t)$.\\ 

On the other hand, let us notice  that the adjoint pair of the
Schr\"{o}dinger  equations (1) comes out
\cite{blanch,olk,olk1} as a direct result of an analytic continuation 
in time of the temporally adjoint parabolic problem (call it Euclidean 
\cite{zambr1,risken}):
$$\partial _t\Theta _*(x,t)=D\triangle \Theta _*(x,t) - 
{1\over {2mD}}V(x)\Theta _*(x,t) $$
$${\partial _t\Theta (x,t)=-D\triangle \Theta (x,t) + {1\over {2mD}}
V(x) \Theta (x,t)}\eqno (6)$$
with a suitable (the same as in (1)) potential function $V(x)$,
defining a 
holomorphic semigroup $exp(-tH/2mD), t\geq 0$, and thus a consistent 
system of solutions  $\Theta _*(x,t),\Theta (x,t)$,
which gives rise to the probability measure
with the quantally factorized density $(\Theta \Theta _*)(x,t)$, on
all  (finite) time intervals run by the time parameter $t$.
Its relevance for the standard nonequilibrium statistical physics
processes, we have discussed elswhere, \cite{blanch,olk2}.

An \it indirect \rm effect of the analytic continuation is the
mapping  of the parabolic system (6) into rather complicated
(nonlinear coupling) parabolic system (5) which is a mathematical,
eventually probabilistic,
equivalent of the Schr\"{o}dinger equation. The seemingly strange 
form of (5) does not preclude the full-fledged stochastic analysis.
In fact, 
the standard methods appropriate for the problem (6) need only 
a slight generalization  to encompass the  time-dependent potentials, 
and next some boundary data analysis to deal with the (a priori
admitted 
by (1)) nodal surfaces of the probability distribution, see e.g. 
\cite{simon,zambr,carm,blanch,freid,gol}. Albeit, in case of (5)
and  (6),  if (1) is not invoked at all,
the time adjoint parabolic equation might
look annoying for the reader unfamiliar with the properties of
fundamental  solutions of the parabolic equations (assuming their
existence in the present context, cf. \cite{zambr,blanch}).
 Let us stress that there is no conflict with the traditional intuition
 about physically irreversible random transport phenomena.\\

In each of the considered problems (1), (5), (6), the probability
density  was naturally associated with  the temporally adjoint
pair of partial differential equations.
Let us choose a concrete time interval $t\in [0,T]$
and consider the boundary data $\rho (x,0), \rho (x,T)$ of the
respective  probability density,
which we demand to be \it strictly positive \rm
on their domain of definition. 
We are interested in deducing a \it stochastic 
process \rm taking place in this time interval, which either
induces a  continuous  propagation (is measure preserving in the
stationary case)
of a probability density between the 
boundary data, or is consistent with the time 
evolution of $\rho (x,t), t\in [0,T]$, if given a priori  as in case
of (1) and  (5).

Since the global existence/uniqueness  theorems \cite{jam,carlen,nel}
tell us that the pertinent processes might be Markovian, we fall
into the well established framework, where for any two Borel sets
$A,B\subset R$ on which
the respective strictly positive boundary  densities $\rho (x,0)$  
and $\rho (x,T)$ are defined, the transition probability
$m(A,B)$ from the set $A$ to the set $B$ in
the time interval $T>0$ has a density given in a specific factorized
form: 
$$m(x,y)=f(x)k(x,0,y,T)g(y)$$
$$m(A,B)=\int_Adx\int_Bdy \, m(x,y)$$
$${\int dy m(x,y)=\rho (x,0)\, ,\, \int dx m(x,y)=\rho (y,T)}
\eqno (7)$$

Here, $f(x), g(y)$ are the a priori unknown functions, to come out as 
solutions of the integral (Schr\"{o}dinger) system of equations (7), 
provided that in addition to the density boundary data we have in
hands 
any strictly positive, continuous  in space variables \it function \rm 
$k(x,0,y,T)$. 
Our notation makes explicit the dependence (in general irrelevant) 
 on the time interval endpoints.  It anticipates an important
restriction we shall impose, that $k(x,0,y,T)$ is a particular form of
a strongly continuous dynamical semigroup  kernel:  it will secure  the
Markov property of the sought for stochastic process.

It is the major mathematical discovery \cite{jam,zambr} that
the Schr\"{o}dinger system (7) of integral equations
admits a unique solution in terms of two nonzero, locally integrable 
(i.e. integrable on compact sets) functions $f(x), g(y)$ of the same 
sign (positive, everything is up to a multiplicative constant).

If $k(y,0,x,T)$ is a particular, confined to the time interval
endpoints,  form of a concrete semigroup kernel
$k(y,s,x,t), 0\leq s\leq t<T$, let
it be a fundamental solution associated with (6) (whose existence
a priori is \it not \rm granted),  
then there exists 
\cite{zambr,blanch,olk,olk1,nag} a function $p(y,s,x,t)$:
$${p(y,s,x,t)=k(y,s,x,t){{\theta (x,t)}\over {\theta (y,s)}}}
\eqno (8)$$
where 
$${\theta (x,t)=\int dy k(x,t,y,T)g(y)}\eqno (9)$$
$$\theta _*(y,s)=\int dx k(x,0,y,s)f(x)$$
which implements a consistent propagation of the density 
$\rho (x,t)=\theta (x,t)\theta _*(x,t)$ between its boundary versions,  
according to:
$${\rho (x,t) = \int p(y,s,x,t)\rho (y,s)dy}\eqno (10)$$
$$0\leq s\leq t<T$$
 For a given semigroup which is characterized by its generator
 (Hamiltonian), the kernel $k(y,s,x,t)$ and the emerging transition
 probability density
$p(y,s,x,t)$  are unique in view of the uniqueness of solutions
$f(x),g(y)$  of (7). For Markov processes, the knowledge of the
transition probability density $p(y,s,x,t)$ for all intermediate
times $0\leq s< t\leq T$  suffices
for the derivation of all other relevant characteristics. 

In the framework of the Schr\"{o}dinger problem the choice of the
integral  kernel $k(y,0,x,T)$ is arbitrary,
except for the strict positivity and
continuity demand. As long as there is no "natural" 
physical motivation for its concrete functional form, the problem is  
abstract and of no direct physical relevance. 
However, in the context of parabolic equations (5) and (6),
this "natural" choice is automatically settled if the Feynman-Kac
formula can be utilized to represent solutions.
(Notice that in case of (5) the finite energy 
condition $\int[|\nabla \psi |^2 + V(x)\rho (x,t)]dx < \infty$, 
$ \rho =|\psi |^2$ secures the boundedness from below of the
potential $2Q(x,t)-V(x)$, see e.g. \cite{zambr,nel,carlen}).

Indeed, in this case an unambigous strictly positive 
semigroup kernel which is a continuous function of its arguments,
can be 
introduced for a broad class of (admissible \cite{simon})
potentials. Time 
dependent potentials are here included as well.
Moreover, in Ref. \cite{blanch} we have discussed a possible 
phenomenological significance  of the Feynman-Kac potentials,
as contrasted  
to the usual identification of Smoluchowski drifts with force fields 
affecting particles (up to a coefficient) in the standard theory of 
stochastic diffusion  processes.\\ 

 There is an enormous literature on the Kac  integral kernel
issue \cite{glimm,simon,freid} based on the concept of the conditional
Wiener measure. Let us however mention that strictly positive
semigroup  kernels generated by
Laplacians plus suitable potentials are very special examples in 
a surprisingly rich encompassing family. The concept of the 
"free noise", normally characterized by a Gaussian probability 
distribution appropriate to a Wiener process, can be extended to 
all infinitely divisible probability distributions via, the well 
known to probabilists and mathematical physicists, L\'{e}vy-Khintchine 
formula, see for example \cite{olk1}.
It allows to expand the  framework from continuous diffusion processes 
to  jump or combined diffusion--jump propagation scenarios which are
not  necessarily Gaussian. All such (L\'{e}vy) processes
are associated with the strictly positive dynamical semigroup kernels
and  the same pertains to a number of cases when the free generator
(minus Laplacian in the "normal" situation) acquires
a potential term, to form a nontrivial Hamiltonian of a physical
problem.\\ 

In the existing probabilistic investigations 
\cite{zambr,zambr1,blanch,olk}, based on the exploitation of the
Schr\"{o}dinger problem strategy, much stronger demand than any
previous  one was in use: guided by
the observation that $k(y,s,x,t)$ must be a \it function \rm  to
allow  for all advantages of (7), it was generally assumed that
the kernel actually  \it is  \rm  a  fundamental solution of the
parabolic equation.
It means that the kernel is a function with continuous 
derivatives: first order-with respect to time, second order-with
respect to space variables.
Then, the transition probability density defined by (8) is a
fundamental solution of the Fokker-Planck (second Kolmogorov)
 equation in the pair $x,t$ of variables, and as such is at the
 same time a solution of the
backward (first Kolmogorov) equation in the pair $y,s$.
This feature was  exploited in \cite{zambr,blanch,olk}.

There is a number of mathematical subtleties involved in the
fundamental solution notion,
since in this case, the Feynman-Kac kernel must be a continuously 
differentiable function, and a solution of  the parabolic equation
itself.
In fact, for suitable (not too bad) potentials, each fundamental 
solution of the parabolic equation has the Feynman-Kac representation, 
\cite{freid}, and is both strictly positive and continuous 
integral kernel \cite{glimm,simon}. 
The inverse statement is generally incorrect: Feynman-Kac kernels may 
have granted the existence status, even as continuous functions
\cite{simon,simon1}, but may not  be differentiable,
and need not to be solutions of any conceivable partial differential 
equations. 
Even, if the Feynman-Kac path integral representation 
applies to explicit solutions of the parabolic equations, 
which are generated from the smooth initial data by the strongly
continuous semigroup action  of the type  $[exp(-tH/2mD)f](x)=
\theta _*(x,t)$, compare e.g. Eq.(9).

To our knowledge, this complication in the study of Markovian 
representations of the Schr\"{o}dinger  interpolating dynamics
(and the quantum Schr\"{o}dinger picture dynamics in particular)
for the first time has  been addressed and solved in Ref. \cite{olk2}.

As well, the subject of the (continuous) 
differentiability of Feynman-Kac kernels seems to have been left aside, 
also in the specialized monographs \cite{glimm,simon,freid}.
Nevertheless, we can firmly repeat the conclusion of our previous
paper \cite{olk} that to give a definite (unique) Markov solution 
of the Schr\"{o}dinger stochastic interpolation problem, in particular 
for the  case of the Schr\"{o}dinger picture 
quantum dynamics (1), a suitable (compatible with (5)) Feynman-Kac 
semigroup with its strictly positive and continuous in all 
variables  kernel \it must \rm  be singled out.  
As it appears, the kernel may  not be a fundamental 
solution of a parabolic equation. Anyway, for each chosen kernel, 
the associated Markov process is defined uniquely by (7)-(8), though 
 \it not \rm in reverse. 

\section{When diffusion processes ?}

The strategy  of deducing a probabilistic solution of the
Schr\"{o}dinger boundary data problem in terms of Markov stochastic
processes running in  continuous time, was accomplished in Ref.
\cite{olk2} in a number
of steps accompanied by the gradual strengthening of restrictions
imposed on the Feynman-Kac potential.
To deal with the commonly accepted diffusion process notion,
 certain additional restrictions need to be imposed to guarrantee
that the mean and variance of the infinitesimal displacements of the
continuous process  have the standard meaning of the drift and diffusion
coefficient, respectively, \cite{horst}.

According to the general wisdom,  diffusions arise in conjunction
with the parabolic evolution equations, since then only the
conditional averages are believed to make sense in the local
description of the
dynamics. It is not accidental that forward parabolic equations
(6) are commonly called the generalized diffusion equations. Also,
the fact that the Feynman-Kac formula involves the integration over
sample paths of the Wiener process, seems to  suggest  some diffusive
features of the Schr\"{o}dinger interpolation, even if we are unable
to establish this fact in a canonical manner.

Clearly, the conditions valid for any $\epsilon >0$: \\
(a) there holds
$lim_{t\downarrow s}{1\over
{t-s}}\int_{|y-x|>\epsilon } p(y,s,x,t)dx=0$,
(notice that (a) is a direct consequence of the stronger, Dynkin
condition, (34)),\\
(b) there exists a drift  function
$b(x,s)=lim_{t\downarrow s}{1\over {t-s}}\int_{|y-s|
\leq \epsilon }(y-x)p(x,s,y,t)dy$, \\
(c) there exists  a diffusion function
$a(x,s)=lim_{t\downarrow s}{1\over {t-s}}
\int_{|y-x|\leq \epsilon }  (y-x)^2 p(x,s,y,t)dy$,\\
are conventionally interpreted to define a diffusion process,
\cite{horst}.

If we exploit  the propagation formula for $\rho (x,t)$,
(10) and ask for the  circumstances under which $\rho (x,t)$ is a
solution of a suitable (Fokker-Planck) parabolic differential
equation, it appears (see e.g. chap. 4.4 in Ref.\cite{horst}),
that the above  conditions (a), (b), (c)
appear to be sufficient but \it not \rm necessary to achieve
this goal. Obviously, they can be satisfied if $p(y,s,x,t)$ is a
fundamental solution.\\

 Usually, one accepts that sample paths of the Wiener
process are continuous with probability one and makes a
\it kinematical \rm 
assumption, \cite{nel,nel1}, (proposal, according to \cite{carlen})
by considering only these processes with continuous trajectories
which can be derived by suitable modifications of the Wiener noise,
and thus are regarded as being of diffusive type from the beginning.
It is
at this point, where the modern theory of stochastic differential 
equations and related probability measures intervenes,
\cite{freid,carm,blanch,carlen}.  Then, an absolute continuity of
measures
relative to the Wiener one, allows for a continuous  (and eventually
diffusion process) realization of the Schr\"{o}dinger interpolation
problem. It arises  in terms of weak (since an initial probability
density $\rho _0(x)$ is attributed to the random variable $X(t)$)
solutions $X(t)=\int_0^t b(X(s),s)ds  +\sqrt{2}\: W(t)$
of respective stochastic differential equations. Here, $W(t)$ stands
for the standard Wiener noise, and $b(x,s)$ is a forward drift of
the diffusion process. Rules of the stochastic It\^{o} calculus allow
to deduce the partial differential (Fokker Planck or second
Kolmogorov) equation governing the dynamics of
the probability density (and of the transition density in particular)
associated with  the process. \\

Since, in the present framework, the Feynman-Kac semigroup kernel and
the related parabolic equations (5), (6) are the principal building blocks
for all the derivations, it seems instructive to indicate  the
standard procedures,\cite{horst,zambr}, linking parabolic
equations with diffusion processes. All of them are based on the
exploitation of fundamental solutions.

We take for granted  the Feynman-Kac representation
 of the continuous and strictly positive kernel associated with
the
forward parabolic equation. Let us assume that we have given a
bounded solution $u(x,t)=\int k(y,0,x,t)u(y,0)dz$ of (5) or (6).
Let us consider $u(x,t)$ and $u(x,t+\triangle t)$, $0\leq \triangle t
\ll 1$.
Since $u(x,t)$ is a solution, we have granted the existence of
the time derivative and the validity of Taylor series with respect
to $\triangle s$, at least to the second expansion order.
The same (at least to third expansion order) applies
to the Taylor expansion of $u(y,t)=u(x+(y-x),t)$ about $x$.
Consequently:
$${u(x,t+\triangle t)=\int k(y,t,x,t+\triangle t)u(y,s)dy
\simeq }\eqno (11)$$
$$\int k(y,t,x,t+\triangle t)[u(x,t)+(y-x)\nabla _xu(x,t) +
{1\over {2!}}
(y-x)^2\triangle _xu(x,t)+...]dy$$
On the other hand, we have
$${u(x,t+\triangle t)\simeq u(x,t)+\partial _tu(x,t)\triangle t}
\eqno (12)$$
and an obvious expansion, in terms of moments of the kernel
$k(y,s,x,t)$ does emerge:
$${\partial _tu(x,t)\triangle t\simeq u(x,t+\triangle t)-u(x,t)
\simeq
-u(x,t)[1-\int k(y,t,x,t+\triangle t)dy] + }\eqno (13)$$
$$[\nabla _xu(x,t)] \int (y-x)k(y,s,x,t+
\triangle t)dy + [{1\over {2!}}\triangle _xu(x,t)]
\int (y-x)^2 k(y,t,x,t+\triangle t)dy+ ... $$
Clearly, to reconcile this expansion with the forward parabolic
equation obeyed by $u(x,t)$, i.e. $\partial _tu=-cu + \triangle u$,
one  needs to verify whether the correct
limiting properties are respected by the Feynman-Kac kernel.   In case
they would hold true, the arguments of Ref. \cite{zambr} would
convince us that we are dealing with the diffusion process.

Presently, \cite{olk2}, the
rigorous demonstration is available in case, when the kernel is
\it not \rm a fundamental solution of the parabolic equation.

\section{From positive to nonnegative solutions
of para\-bolic equations}

Following Refs. \cite{olk2,jmp1},
let us  focus our attention  on stochastic Markov
processes of diffusion-type (see Ref. \cite{olk1} for a jump process
alternative), which are associated with the general temporally
adjoint pair of parabolic partial differential equations:
$${\partial _tu(x,t)=\triangle u(x,t)-c(x,t)u(x,t)}\eqno (14)$$
$$\partial _tv(x,t)=-\triangle v(x,t)+c(x,t)v(x,t)$$
Here, $c(x,t)$ is a real function (left unspecified at the moment)
and the solutions $u(x,t)$, $v(x,t)$ are sought for in the time
interval $[0,T]$ under the boundary conditions set at the
time-interval borders:
$${\rho _0(x)=u(x,0)v(x,0)}\eqno (15)$$
$$\rho _T(x)=u(x,T)v(x,T)$$
$$\int_A\rho _0(x)dx=\rho _0(A)\; ,\;
\int_B\rho _T(x)dx=\rho _T(B)$$
We assume that $\rho $ is a probability measure with the density
$\rho (x)$, and $A,B$ stand for arbitrary Borel sets in the
event space. In the above, suitable units were chosen to eliminate
inessential in the present context (dimensional) parameters, and
the process is supposed to live in/on $R^1$.

As emphasized in the previous publications, \cite{blanch}-\cite{olk2},
the key ingredient of the formalism is to specify the function
$c(x,t)$
such that $exp[-\int_0^tH(\tau )d\tau ]$ 
can be viewed as a strongly continuous semigroup operator
with the generator   $H(t)=-\triangle +c(t)$, associated with the
familiar \cite{simon} Feynman-Kac kernel:
$${(f,exp[-\int_0^tH(\tau )d\tau ]g)=\int dy \int dx
\overline{f}(y)k(y,0,x,t)g(x)=}\eqno (16)$$
$$\int \overline{f}(\omega (0)) g(\omega (t)) exp[-\int_0^tc(\omega
(\tau ),\tau )d\tau ] d\mu _0(\omega )$$
The exponential operator should be understood as the time-ordered
expression.
Here $f,g$ are complex functions, $\omega (t)$ denotes a sample path
of the conventional Wiener process and $d\mu _0$ stands for the Wiener
measure. Clearly, the kernel itself  can be explicitly written
in terms
of the conditional Wiener measure $d\mu _{(x,t)}^{(y,s)}$ pinned at
space-time points $(y,s)$ and $(x,t)$, $0\leq s<t\leq T$:
$${k(y,s,x,t)=\int exp[-\int_s^t c(\omega (\tau ),\tau )d\tau ]\: d\mu
^{(y,s)}_{(x,t)}(\omega )}\eqno (17)$$
As long as we do not impose any specific domain restrictions on the
semigroup generator $H(\tau )$, the whole real line $R^1$ is
accessible to the process.
Various choices of the Dirichlet \cite{simon,blanch}
boundary conditions can be accounted for  by the formula (3). If we
replace $R^1$ by  any open subset $\Omega \subset R^1$ with the
boundary $\partial \Omega $,
it  amounts to confining  Wiener sample paths of
relevance to reside in  (be interior to) $\Omega $,
which in turn needs an appropriate measure
$d\mu _{(x,t)}^{(y,s)}(\omega \in \Omega )$ in (4).
This is usually implemented by means of stopping times for the
Wiener process, \cite{nag,blanch,carm,gol}.

Let $f(x)$, $g(x)$ be two real functions such that:
${m_T(x,y)=f(x)k(x,0,y,T)g(y)}$
defines a bi-variate density of the probability measure,
i.e. a transition probability of the propagation from the Borel set
$A$ to the Borel set $B$ to be accomplished in the time interval $T$.
In particular, we need the marginal probability densities
 to be defined:
${\rho_0(x)=m_T(x,\Omega )\; , \; \rho _T(y) =m_T(\Omega ,y)}$
where $\Omega \subset R^1$ is a spatial  area confining  the process.

These formulas  can be viewed as special cases  of (7), so
establishing an apparent link between the Schr\"{o}dinger
problem and  the Feynman-Kac  kernels, together with the
related parabolic equations.
Assuming that marginal probability  measures  and their densities
are given a priori, and a concrete  Feynman-Kac kernel (17)
(with or without Dirichlet  domain restrictions) is specified,
we are within the premises of the Schr\"{o}dinger
boundary data problem.

Let $\overline{\Omega }=\Omega \cup \partial \Omega $ be a closed
subset of $R^1$, or $R^1$ itself. For all Borel sets (in the $\sigma
$-field generated by all open subsets of $\overline{\Omega }$) we
assume to  have known $\rho _0(A)$ and $\rho _T(B)$, hence the
respective densities as well. If the integral kernel $k(x,0,y,T)$
in the expression (5) is chosen to be \it continuous \rm and \it
strictly positive \rm on $\overline{\Omega }$, then the integral
equations (7) can be solved \cite{jam} with respect to the \it
unknown  \rm functions $f(x)$ and $g(y)$.  The solution comprises
two nonzero, locally integrable functions  of the same sign, which
are unique up to a  multiplicative constant.

If, in addition, the kernel $k(y,s,x,t)$, $0\leq s<t\leq T$ is a \it
fundamental solution \rm  of the parabolic system (14) on
$R^1$ (i.e. is a function which solves the forward equation in $(x,t)$
variables, while the backward one in $(y,s)$), then we have defined a
 solution of the system (14) by:
$${u(x,t)\equiv f(x,t)=\int f(y) k(y,0,x,t)dy}\eqno (18)$$
$$v(x,t)\equiv g(x,t)=\int k(x,t,y,T)g(y)dy$$
Moreover, $\rho (x,t)=f(x,t)g(x,t)$ is propagated by the Markovian
transition probability density:
$${p(y,s,x,t)=k(y,s,x,t){{g(x,t)}\over {g(y,s)}}}\eqno (19)$$
$$\rho (x,t)=\int \rho (y,s) p(y,s,x,t) dy $$
$$0\leq s<t\leq T$$
$$\partial _t\rho =\triangle \rho - \nabla (b\rho )$$
$$b=b(x,t)=2{{\nabla g(x,t)}\over g(x,t)}$$
the result, which covers all traditional Smoluchowski diffusions
\cite{blanch,garb}. In that case, $c(x,t)$ is regarded as
time-independent, and the corresponding stochastic process is
homogeneous in time.
The Dirichlet boundary data can be implemented as well,
thus leading to the Smoluchowski diffusion processes with natural
boundaries, \cite{blanch}.
Then, $k(y,s,x,t)$ stands for an appropriate
Green function of the parabolic boundary-data problem,
with the property
to vanish at the boundaries $\partial \Omega $ of $\Omega $.

Let us  mention that for time-independent potentials,
$c(x,t)=c(x)$ for all $t\in [0,T]$, a number of generalizations is
available \cite{nag,blanch,carm,combe,gol}-\cite{fuk} to
encompass
the nodal sets of $\rho (x)$ and hence of the associated functions
$f(x),g(x)$.  The drift
$b(x)=\nabla ln\rho (x)={{\nabla \rho (x)}\over {\rho (x)}}$
singularities  do not prohibit the existence of a well defined
Markov diffusion process (9), for which  nodes are unattainable.
In the considered
framework  they are allowed only at the boundaries of the
connected spatial area $\Omega $ confinig the process.

The problem of relaxing the strict positivity (and/or continuity)
demand for Feynman-Kac kernels is nontrivial \cite{fort,beur,jam}
with respect to the eventual construction of the \it unique \rm
Markov process (9). To elucidate the nature of difficulties
underlying
this issue, we shall  consider  quantally motivated examples of the
parabolic dynamics (14).

\section{Nonlinear parabolic dynamics with unattainable boundaries}

Let us choose the potential function $c(x,t)$ as follows:
$${c(x,t)={x^2\over {2(1+t^2)^2}} - {1\over {1+t^2}}}\eqno (20)$$
for $x\in R^1, t\in [0,T]$.
In view of its local H\"{o}lder continuity
(cf. Ref. \cite{olk2}) with exponent one, and its quadratic
boundedness, the
fundamental solution of the parabolic system  is known to exist
\cite{freid,kal,szyb,bes}. It is constructed via the
 parametrix method. Among an
infinity of regular solutions of (14) with the potential (20), we can
in particular identify \cite{olk2} solutions of the Schr\"{o}dinger
boundary data problem for the familiar (quantal) evolution:
$${\rho _0(x)=(2\pi )^{-1/2} exp[-{x^2\over 2}] \longrightarrow
\rho (x,t)=[2\pi (1+t^2)]^{-1/2} exp[-{x^2\over {2(1+t^2)}}]}\eqno
(21)$$
They read
$${u(x,t)\equiv f(x,t)=[2\pi (1+t^2)]^{-1/4} exp(-{x^2\over
4}{{1+t}\over {1+t^2}}+ {1\over 2} arctan\: t)}\eqno (22)$$
$$v(x,t)\equiv g(x,t)=[2\pi (1+t^2)]^{-1/4} exp(-{x^2\over 4}{{1-t}
\over {1+t^2}}- {1\over 2} arctan\: t)$$
and, while solving the nonlinear parabolic system (14) (with
$c=\triangle \rho ^{1/2}/\rho ^{1/2}$), in addition they imply the
validity of the Fokker-Planck equation:
$$\rho (x,t)= f(x,t)g(x,t) \rightarrow \partial _t\rho =
\triangle \rho - \nabla (b\rho )$$
$${b(x,t)=2{{\nabla g(x,t)}\over {g(x,t)}} =
- {{1-t}\over {1+t^2}}\: x}\eqno (23)$$
Notice that $p(y,s,x,t)=k(y,s,x,t){{g(x,t)}\over {g(y,s)}}$ is  a
fundamental solution of the first and second Kolmogorov (e.g.
Fokker-Planck) equations in the present case.

Let us recall that a concrete parabolic system corresponding to
solutions (22) looks badly nonlinear.
Our procedure, of first considering
the linear system (but with the potential "belonging" to another,
nonlinear one), and next identifying solutions of interest by means of
the Schr\"{o}dinger boundary data problem,  allows to bypass this
inherent difficulty.
In connection with the  previously mentioned quantal motivation of
ours,
let us define $g=exp(R+S), f=exp(R-S)$  where $R(x,t), S(x,t)$
are real
functions. We immediately realize that (5), (14) provide for
a parabolic alternative to the familiar Schr\"{o}dinger equation
and its temporal adjoint:
$${i\partial _t\psi = -\triangle \psi }\eqno (24)$$
$$i\partial _t\overline{\psi } = \triangle \overline{\psi }$$
with the Madelung factorization $\psi =exp(R+iS)$, $\overline{\psi
}=exp(R-iS)$ involving the previously introduced real functions $R$
and $S$.

Things seem to be fairly transparent when the parabolic system (5)
 or (6)
allows  for  fundamental solutions. However, even in this case
complications arise if nodes of the probability density are admitted.
The subsequent discussion has a quantal origin,
and comes from  the free Schr\"{o}dinger propagation  with the
specific choice, \cite{jmp1},  of the initial data:
$${\psi _0(x)=(2\pi )^{-1/4} \: x\: exp(-{x^2\over 4})
\longrightarrow
}\eqno (25)$$
$$\psi (x,t)=(2\pi )^{-1/4} {x\over {(1+it)^{3/2}}}\: exp[-{x^2\over
{4(1+it)}}]$$
such  that  our nonstationary dynamics example displays a stable
node at $x=0$ for all times .

The  parabolic system (1) in this case involves the potential
function:
$${c(x,t)={{\triangle \rho ^{1/2}(x,t)}\over {\rho ^{1/2}(x,t)}}=
{x^2\over {2(1+t^2)^2}}- {3\over {1+t^2}}}\eqno (26)$$
$$\rho (x,t)=(2\pi )^{-1/2}(1+t^2)^{-3/2}\: x^2\: exp[-{x^2\over
{2(1+t^2)}}]$$
The polar (Madelung) factorization of Schr\"{o}dinger wave functions
implies:
$${R(x,t)=ln\: \rho ^{1/2} (x,t)}\eqno (27)$$
$$x>0\rightarrow S(x,t)=S_+(x,t)=
{x^2\over 4}{t\over {1+t^2}} -{3\over 2}arctan\: t
$$
$$x<0\rightarrow S(x,t)=S_-(x,t)=
{x^2\over 4}{t\over {1+t^2}}-{3\over 2}arctan \: t \:
+\: \pi $$
Although $S(x,t)$ is not defined at $x=0$, we can introduce continuous
functions $f=exp(R-S)$ and $g=exp(R+S)$ by employing the step function
$\epsilon (x)=0$ if $x\geq 0$ and $\epsilon (x)=1$ if $x<0$. Then, the
candidates for solutions of the parabolic system (5) with the
potential (20) would read:
$${v(x,t)\equiv g(x,t)=(2\pi )^{-1/4} (1+t^2)^{-3/4} |x|\:
exp(-{x^2\over 4}{{1-t}\over {1+t^2}})\:
exp[-{3\over 2}arctan\: t + \pi
\epsilon (x)]}\eqno (28)$$
$$u(x,t)\equiv f(x,t)= (2\pi )^{-1/4}(1+t^2)^{-3/4} |x|\:
exp(-{x^2\over
4}{{1+t}\over {1+t^2}})\:
exp[{3\over 2}arctan\: t - \pi \epsilon (x)]$$
For all $x\neq 0$ we can define the forward drift
$${b(x,t)=2{{\nabla g(x,t)}\over {g(x,t)}}={2\over x} - x\: {{1-t}
\over
{1+t^2}}}\eqno (29)$$
which displays a singularity at $x=0$.
Nonetheless, $(b\rho )(x,t)$ is a smooth function and the
Fokker-Planck
equation $\partial _t\rho =\triangle \rho - \nabla (b\rho ) $
holds true
on the whole real line $R^1$, for all  $t\in [0,T]$.
Notice that there is no current through $x=0$, since $v(x,t)=2\nabla
S(x,t)={{xt}\over {1+t^2}}$ vanishes at this point for all times.

Our functions $f(x,t), g(x,t)$ are continuous on $R^1$, which  however
does not imply their differentiability. Indeed, they solve the
parabolic
system (5) with the potential (20) \it not \rm on $R^1$  but on
$(-\infty ,0)\cup (0,+\infty )$. Hence, almost everywhere on $R^1$,
with the exception of $x=0$.

An apparent obstacle arises because of this subtlety: these functions
are \it not \rm even weak solutions of (1), because of:
$${\int_{-\infty }^{+\infty } \partial _tf(x,t)\phi (x) dx +
\int_{-\infty }^{+\infty }\nabla f(x,t) \nabla \phi (x) dx +}\eqno
(30)$$
$${1\over 2} \int_{-\infty }^{+\infty }c(x,t)f(x,t)\phi (x) dx \:
\neq 0
$$
for every test function $\phi $ such that $\phi (0)\neq 0$,
continuous and with support on a chosen compact set (e.g. vanishing
beyond this set).

One more obstacle arises, if we notice that $c(x,t)$, (20) permits the
existence of the unique, bounded and strictly positive fundamental
solution   for the parabolic system (14). Then, while having singled
out
a fundamental solution and the boundary density data
$\rho _0(x), \rho
_T(x)$ consistent with (20), we can address the Schr\"{o}dinger
boundary data problem associated with (2),(3):
$${u(x,0)\int k(x,0,y,T)v(y,T)dy = \rho _0(x)}\eqno (31)$$
$$v(x,T)\int k(y,0,x,T)u(y,0)dy =\rho _T(x)$$
expecting that a unique solution $u(x,0),v(x,T)$ of this system of
equations implies an identification $u(x,0)=f(x,0)$ and
$v(x,T)=g(x,T)$.

However, it is not the case and our $f(x,t),g(x,t)$ do not
come out as
solutions of the Schr\"{o}dinger problem,
if considered on the whole real line $R^1$,
on which the fundamental solution sets rules of the game.
Indeed, let us assume that (31) does hold true if we choose
$u(x,0)=f(x,0),
v(x,T)=g(x,T)$, with $f$ and $g$ defined by (21). Since,
in particular we have
$${g(x,T)\int k(y,0,x,T)f(y,0)dy = g(x,T)f(x,T)}\eqno (32)$$
then for $x\neq 0$ there holds:
$${f(x,T)=\int k(y,0,x,T)f(y,0)dy}\eqno (33)$$
Both sides of the last identity represent continuous functions, hence
the equality is valid point-wise (i.e. for every $x$). We know that
$f(y,0)$ is continuous and bounded on $R^1$, and $k(y,0,x,T)$ is a
fundamental solution of (1). Hence the right-hand-side of (33)
represents a regular solution of the parabolic equation.
Such solutions
have continuous derivatives, while our left-hand-side function
$f(x,T)$ certainly does not share this property.
Consequently, our assumption
leads to a contradiction and (33) is invalid in our case.

It means that the fundamental solution (e.g. the corresponding
Feynman-Kac kernel) associated with (16) is inappropriate for the
Schr\"{o}dinger problem analysis, if the interpolating probability
density is to have nodes (i.e. vanish at some points).

In our case, $x=0$ is a stable node of $\rho (x,t)$, and is a
time-independent repulsive obstacle for the stochastic process. An
apparent way out of the situation comes by considering two
non-communicating processes, which are separated by the unattainable
barrier at $x=0$, \cite{gol,carlen,blanch,alb}.  The pertinent
discussion can be found in \cite{jmp1}.

\section{The "Wiener exclusion"}

The conventional definition of the Feynman-Kac kernel (in the
conservative case)
$${exp[-t(-\triangle +c)](y,x)=\int exp[-\int_0^t
c(\omega (\tau ))d\tau ] d\mu _{(x,t)}^{(y,0)} (\omega )}\eqno (34)$$
comprises all sample paths of the Wiener process on $R^1$, providing
merely for their nontrivial  redistribution by means of the
Feynman-Kac weight $exp[-\int_0^tc(\omega (\tau ))d\tau ]$ assigned to
each sample path $\omega (s): \omega (0)=y, \omega (t)=x$.

Assume that $c(x)$ is bounded from below and locally (i.e. on compact
sets) bounded from above. Then, the kernel is strictly positive and
 continuous \cite{simon}.

For $c=0$ we deal with the conditional Wiener measure
$${exp(t\triangle )(y,x)=\mu _{(x,t)}^{(y,0)} [\omega (s)\in R^1;
0\leq s\leq t]=}\eqno (35)$$
$$\mu[\omega (s)\in R^1;\omega (0)=y,
\omega (t)=x; 0\leq s\leq t]$$
pinned at space-time points $(y,0)$ and $(x,t)$.

The previous discussion indicates that $R_-$ is
inaccessible for all sample paths originating from $R_+$.
In reverse, $R_+$ is inaccessible for those from $R_-$.
As well, we may confine the process to an arbitrary closed subset
$\Omega \subset R^1$, or enforce it to avoid ("Wiener exclusion" of
Ref. \cite{simon1}) certain areas in $R^1$.

In this context, it is instructive to know that \cite{simon} for an
arbitrary open set $\Omega $, there holds:
$${exp(t\triangle _{\Omega }) (y,x)= \mu _{(x,t)}^{(y,0)}
[\omega (s)\in
\Omega , 0\leq s\leq t]}\eqno (36)$$
which is at the same time  a  definition of  the operator
$-\triangle _{\Omega }$,
i.e. the Laplacian with Dirichlet boundary conditions, and that of
the associated semigroup kernel. This formula provides us with the
conditional Wiener measure which is \it confined \rm to the interior
of a given open set,\cite{gin,brat,blanch}.

We can introduce an analogous measure, which is confined to the \it
exterior \rm of a given closed subset $S\subset R^1$.
In case of not
too bad sets (like an exterior of an interval  in $R^1$ or a ball in
$R^n$, the corresponding integral kernel in known \cite{simon} to be
positive and continuous. Technically, if $S$ is a (regular) closed set
such that the Lebesgue measure of $\partial S$ is zero, then:
$${exp(t\triangle _{R\backslash S})(y,x)=\mu _{(x,t)}^{(y,0)}
[\omega (s)
\notin S; 0\leq s\leq t]}\eqno (37)$$

The Feynman-Kac spatial redistribution of Brownian paths  can
be extended to cases (36),
(37) through the general formula valid for any $f,g \in L^2(\Omega )$,
where $\Omega $ is any open set of interest (hence $R\backslash S$, in
particular):
$${(f,exp(-tH_{\Omega } )g)=\int_{\Omega } \overline {f}(\omega (0))
g(\omega (t))exp[-\int_0^tc(\omega (\tau ))d\tau ] d\mu _0(\omega )}
\eqno (38)$$
It gives rise to the integral kernel comprising the restricted Wiener
path integration, which is defined at least almost everywhere in
$x,y$.
Then, its continuity is not automatically granted.
We can also utilize
a concept of the first exit time $T_{\Omega }$ for the sample path
started inside $\Omega $ (or outside $S$)
$${T_{\Omega }(\omega )=inf[t>0, X_t(\omega )\notin \Omega ]}\eqno (39)$$
where $X_t$ is the random variable of the process. Then, we can write,
\cite{carm,blanch}
$${exp(-tH_{\Omega })(y,x)=\int exp[-\int_0^tc(\omega (\tau ))d\tau ]
d\mu _{(x,t)}^{(y,0)} [\omega ; t<T_{\Omega }]=}\eqno (40)$$
$$\int_{\Omega }exp[-\int_0^tc(\omega (\tau ))d\tau ]d\mu
_{(x,t)}^{(y,0)}(\omega )$$
It is an integration restricted to these Brownian paths, which while
originating from $y\in \Omega $ at time $t=0$ are conditioned to
reach $x\in \Omega $ at time $t>0$ without crossing (but possibly
touching) the boundary $\partial S$ of $S$.
The contribution from paths which would touch the boundary without
crossing, for at least one instant $s\in [0,t]$ is of Wiener measure
zero, \cite{gin}.

In case of processes with unattainable boundaries,  with
probability 1, there is no sample path which could possibly
reach the barrier at any instant $s<\infty $.

The above discussion made an implicit use of the integrability
property
$${\int_0^tc(\omega (s))ds <\infty}\eqno (41)$$
for $\omega \in R^1, 0\leq s\leq t$., in which case
the corresponding
integral kernel (for bounded from below potentials) is strictly
positive.
Then, if certain areas are inaccessible to the process, it occurs
excusively \cite{combe}-\cite{fuk} due to the drift singularities,
which are capable of "pushing" the sample paths away
from the barriers.

The previous procedure can be extended to the singular
\cite{faris}-\cite{klaud3} potentials, which are allowed to diverge.
Their study was in part motivated by the so called
Klauder's phenomenon
(and the related issue of the ground state degeneracy of quantal
Hamiltonians), and had received a considerable attention in the
literature.

In principle, if $S$ is a closed set in $R^1$ like before, and
$c(x)<\infty $ for all $x\in \Omega =R\backslash S$,
while $c(x)=\infty
$ for $x\in S$, then depending on how severe the singularity is,
we can
formulate a  criterion to grant the exclusion of certain sample
paths of the process and hence to limit an
availability of certain spatial areas to the random motion.
Namely, in case of (41) nothing specific happens, but if we have
$${\int_0^t c(\omega (\tau ))d\tau =\infty }\eqno (42)$$
for $\omega (\tau ) \in S$ for some $\tau \in [0,t]$, then the
"Wiener exclusion" certainly appears: we are left with
contributions from these sample paths only for which (42) does not
occur.  Unless the respective set is of Wiener measure zero.

The area $\Omega $ comprising  the relevant sample paths is then
selected as follows:
$${\Omega =[\omega ; \int_0^tc(\omega (\tau ))d\tau < \infty ]}\eqno
(43)$$

In particular, the criterion (43) excludes from considerations sample
paths which cross $S$ and so would establish a communication
between the distinct connected components of $\Omega $.

The singular set $S$ can be chosen to be of Lebesgue measure zero and
contain a finite set of points dividing $R$ into a finite number
of open
connected components. With each open and connected subset $\Omega
\subset R^1$ we can \cite{faris} associate a strictly positive
Feynman-Kac kernel, which can be expected to display continuity.

Since the respective potentials diverge on $S$, their behaviour in a
close neighbourhood of nodes is quite indicative. For, if $\omega _S$
is a Wiener process sample path which is bound to cross  a node at
$0\leq s\leq t$, then the
corresponding contribution to the path integral vanishes.  Such paths
are thus excluded from consideration.
If their subset is sizable (of nonzero Wiener measure), then the
eliminated contribution
$${\int_{\omega _{S}} exp[-\int_0^tc(\omega _S(\tau ))d\tau] d\mu
_{(x,t)}^{(y,0)}(\omega )= 0 }\eqno (44)$$
is substantial in the general formula (35).

At the same time, we get involved a nontrivial domain property of
the semigroup generator $H=-\triangle +c$ resulting in the so called
ground state degeneracy \cite{faris,klaud,klaud1}.
Let us recall (Theorem 25.15 in Ref. \cite{simon})  that if $c$ is
bounded from below and locally bounded from above, then the ground
state function of $H=-\triangle +c$ is everywhere strictly positive
and thus bounded away from zero on every compact set.

\section{Singular potentials,  ground state degeneracy and the
"Wiener exclusion"}

Our further discussion will concentrate mainly on singular
perturbations
of the harmonic potential. Therefore, some basic features of the
respective parabolic problem are worth invoking.
The eigenvalue problem (the temporally adjoint parabolic system
now trivializes):
$${-\triangle g+(x^2-E)g=0=\triangle f-(x^2-E)f}\eqno (45)$$
has well known solutions labeled by $E_n=2n+1$ with $n=0,1,2,...$.
In particular, $g_0(x)=f_0(x)={1\over {\pi ^{1/4}}}exp(-{x^2\over 2})$
is the unique nondegenerate ground state solution. The corresponding
Feynman-Kac kernel reads
$$exp(-tH)(y,x)=k(y,0,x,t)=k_t(y,x)=$$
$${(\pi )^{-1/2} (1-exp(-t))^{-1/2}
exp[-{{x^2-y^2}\over 2} - {{(yexp(-t) -x)^2}\over 2}]}\eqno (46)$$
$$\partial _tk= -\triangle _xk + (x^2-1)k$$
and the invariant probability density $\rho (x)=f(x)g(x)=(\pi
)^{-1/2}exp(-x^2)$ is preserved in the course of the time-homogeneous
diffusion process with the transition probability density
$${p(y,s,x,t)=k_{t-s}(y,x) {{g(x)}\over {g(y)}}}\eqno (47)$$
We have $p(y,s,x,t)=p(y,0,x,t-s)$.

Notice the necessity of the eigenvalue correction (renormalization)
of the potential, both
in (34) and (38), which is indispensable to reconcile the functional
form of the forward drift $b(x)=2\nabla ln g(x)=-2x$ with the general
expression for the corresponding (to the diffusion process) parabolic
system potential
$${c=c(x,t)= \partial _t ln\: g + {1\over 2} ({b^2\over 2} + \nabla
b)}\eqno (48)$$
which equals $c(x)=x^2-1$ in our case.

Let us pass to the  singular (degenerate) problems.
\\
The canonical (in the context of Refs. \cite{faris}-\cite{cal})
choice of the centrifugal potential:
$${c_E(x)=x^2+{{2\gamma }\over x^2} - E}\eqno (49)$$
generates a well known spectral solution \cite{par,cal}
for $Hg=[-\triangle +c_E(x)]g$.  The eigenvalues:
$${E_n=4n+2+(1+8\gamma )^{1/2}}\eqno (50)$$
with $n=0,1,2,...$ and $\gamma >-{1\over 8}\Rightarrow (1+8\gamma
)^{1/2}>\sqrt {2}$, are associated with the eigenfunctions
of the form:
$${g_n(x)=x^{(2\gamma +1)/2}\: exp(-{x^2\over 2})\: L_n^{\alpha
}(x^2)}\eqno (51)$$
$$\alpha =(1+8\gamma )^{1/2}$$
$$L_n^{\alpha }(x^2)=\sum_{\nu =0}^n {{(n+\alpha )!}\over {(n-\nu )!
(\alpha +\nu )!}}\: {{(-x^2)^{\nu }}\over {\nu !}} \longrightarrow $$
$$
L_0^{\alpha }(x^2)=1\; , \; L_1^{\alpha }(x^2)=-x^2+\alpha +1$$
It demonstrates an apparent double degeneracy of both the ground state
and of the whole eigenspace of the generator $H$. The singularity at
$x=0$ does not prevent the definition of $H=-\triangle + x^2 +
{{2\gamma
}\over {x^2}}$ since  this operator is densely defined on an
appropriate subspace of
$L^2(R^1)$. This singularity is sufficiently severe
to
decouple $(-\infty ,0)$ from $(0,\infty )$ so that
$L^2(-\infty ,0)$ and
$L^2(0,\infty )$ are the invariant subspaces of $H$ with the resulting
overall double degeneracy.
\\

Potentials of the form, \cite{simon,simon1,faris}:
$${c(x)=x^2 + [dist(x,\partial \Omega )]^{-3}}\eqno (52)$$
where $\partial \Omega $ can be identified with $\partial S$, and $S$
is a closed subset in $R^1$ of any (zero or nonzero) Lebesgue measure,
have properties generic to the Klauder's phenomenon.
Because the Wiener
paths are known to be H\"{o}lder continuous of any order ${1\over
2}-\epsilon , \epsilon >0$ and of order ${1\over 3}$ in particular,
there holds $\int _0^tc(\omega (\tau ))d\tau =\infty $ if
$\omega (\tau )\in S$ for some $\tau $.
Conversely, $\int _0^t c(\omega (\tau ))d\tau
<\infty $ if $\omega $ never hits $S$.
This implies that the relevant contributions to:
$${(f,exp[-t(-\triangle +c)]g)=\int \overline {f}(\omega (0))
g(\omega
(t)) exp[-\int_0^t c(\omega (\tau ))d\tau ] d\mu _0(\omega )}\eqno
(53)$$
come only from the subset of paths defined by
$${Q_t= [\omega ;\; \int_0^t c(\omega (\tau ))d\tau < \infty
]}\eqno (54)$$
The above argument might seem inapplicable to the centrifugal problem.
However it is not so.  In the discussion of the divergence of certain
integrals of the Wiener process, in the context of Klauder's
phenomenon, it has been proven \cite{klaud2} that for almost
every path from $x=1$
to $x=-1$ (crossing the singularity point $x=0$) there holds
$\int_{t-\delta }^{t+\delta }|\omega (\tau )|^{-1} d\tau = \infty $
for any $\delta >0$.

To be more explicit:
if $\tau _1=\tau _1(\omega )$ is the first time such that the Wiener
process $W(t)=W(t,\omega )$ attains the level (location on $R^1$)
 $W(\tau _1)=1$, then the integral over any right-hand-side
neighbourhood $(\tau _1,\tau _1+\delta )$ of $\tau _1$ diverges:
$${\int_{\tau _1}^{\tau _1+\delta } c(\omega (t)-1)dt=\infty }\eqno
(55)$$
if $\int_{-1}^{+1} c(x)dx = \infty $.\\
In case of the left-hand-neighbourhood of $\tau _1$, we have
$${\int_{\tau _1-\delta }^{\tau _1}c(\omega (t)-1)dt =\infty }\eqno
(56)$$
if $\int_{-1}^0 x c(x)dx =\infty $.

All that holds true in case of the centrifugal potential, thus proving
that the only subset of sample paths, which matters in (54) is (55).
Obviously, $Q_t$ does not include neither paths crossing $x=0$ nor
those which might hit (touch) $x=0$ at any instant.
The singularity is sufficiently severe to create an unattainable
repulsive boundary for all possible processes, which we can associate
with the spectral solution (50),(51).

After the previous analysis one might be left with an  impression that
the appearence of the stable barrier at $x=0$ persisting for all $t\in
[0,T]$, is a consequence of the initial data choice
$\psi _0(0)=0 $  for the involved quantum Schr\"{o}dinger picture
dynamics.  In general it is not so.
For example, $\psi _0(x)=x^2exp(-x^2/4)$ which vanishes at $x=0$, does
not vanish anymore for times $t>0$ of the free evolution.
On the other hand, somewhat surprisingly from the parabolic
(intuition) viewpoint, the node can be dynamically developed from the
nonvanishing initial data and lead to the nonvanishing terminal
data.

Let us consider, \cite{jmp1}, a complex function:
$${\psi (x,t)=(1+it)^{-1/2} exp[-{x^2\over {4(1+it)}}] \; [{x^2\over
{2(1+it)^2}} +{{it}\over {1+it}}]}\eqno (57)$$
which solves the free Schr\"{o}dinger equation with the initial data
$\psi (x,0)={x^2\over 2}exp(-{x^2\over 4})$. It vanishes at $ x=0$
exclusively at the initial instant $t=0$ of the evolution.

Obviously, there is nothing to prevent us from considering
$${\Psi (x,t)=\psi (x,t-\alpha )}\eqno (58)$$
for $\alpha >0$. It solves the same free equation, but with
nonvanishing initial data. However, the node is developed in the
course
of this evolution at time $t=\alpha $ and instantaneously
desintegrated for times $t>\alpha $.
Here, the Schr\"{o}dinger boundary data problem would obviously
involve
two strictly positive probability densities
$\rho _0(x)=|\Psi (x,0)|^2$
and $\rho _T(x)=|\Psi (x,T)|^2,\; T>\alpha $. It would suggest to
utilize the theory \cite{olk2}, based on strictly positive Feynman-Kac
kernels, to analyze the corresponding interpolating process.
However,  this
tool is certainly inappropriate and cannot reproduce the a priori
known
dynamics, with the node arising at the intermediate time instant.

To handle the issue by means of a parabolic system,
which we can always
associate with a quantum Schr\"{o}dinger picture dynamics, let us
evaluate the potential $c(x,t)$ appropriate for (5).

In view of
$${\rho (x,t)= const \; (1+t^2)^{-5/2} exp[-{x^2\over {2(1+t^2)}}]\;
[{x^4\over 4} - x^2t^2 + t^2(1+t^2)]}\eqno (59)$$
we have  (while setting $w^{1/2}(x,t)=[{x^2\over 4}-x^2t^2 +
t^2(1+t^2)]$):
$${c(x,t)={{\triangle \rho ^{1/2}(x,t)}\over {\rho ^{1/2}(x,t)}}=
{1\over
4} (-{x\over {1+t^2}}+\nabla w)^2+
{1\over 2}(-{1\over {1+t^2}}+ \triangle ln \; w)=}\eqno (60)$$
$${1\over 4}{x^2\over {(1+t^2)^2}}-{1\over 2}{{3x^2-2t^2x}\over
{{x^4\over 4} -t^2x^2+ t^2(1+t^2)}}-{1\over {2(1+t^2)}}+$$
$${1\over 2}{{3x^2-2t^2}\over {{x^4\over 4}-t^2x^2+t^2(1+t^2)}}-
{1\over 4}({{x^3-2t^2x}\over {{x^4\over 4} - t^2x^2 +t^2(1+t^2)}})^2$$
The expression looks desparately discouraging, but its $t\downarrow 0$
(i.e. the initial data ) limit is quite familiar and displays a
centrifugal singularity at $x=0$:
$${c(x,t)={{\triangle \rho ^{1/2}(x,0)}\over {\rho
^{1/2}(x,0)}}={x^2\over 4}+ {2\over x^2} - {5\over 2}}\eqno (61)$$

Since the original, dimensional expression for the centrifugal
eigenvalue problem is \cite{par}:
$${(-{1\over 2}\triangle +{m^2\over 2}x^2 + {\gamma \over x^2})g=
Eg}\eqno (62)$$
$$E_n=m[2n+1+{1\over 2}(1+8\gamma )^{1/2}]$$
with $n=0,1,...$, an obvious adjustment of constants $m=1/2,
\gamma =1$
allows to identify $E=5/2$ as the $n=0$ eigenvalue of the centrifugal
Hamiltonian $H=-\triangle +{x^2\over 4} +{2\over x^2}$.

A  peculiarity of the considered example is that it enables us to
achieve an explicit insight into   an emergence of the
centrifugal singularity and its
subsequent destruction (decay) for times $t>\alpha $, due to the free
quantum evolution.

In view of the degeneracy of the ground-state eigenfunction
${x^2\over 2}exp(-{x^2\over 4})$ of the centrifugal Hamiltonian,
we deal here with
the gradually decreasing communication between $R_+$ and $R_-$, which
results in the emergence of the completely separated (disjoint) sets
$(-\infty ,0)$ and $(0,+\infty )$ at $t=\alpha $,  followed by the
gradual increase of the communiaction  for times $t>\alpha $. By
"communication" we understand that the set of sample paths crossing
$x=0$ forms a subset of nonzero Wiener measure.

It also involves a generalisation (cf. also Refs.
\cite{olk2,klaud3,freid}) to  time-dependent Feynman-Kac kernels:
$${(f,exp[-\int_s^tH(\tau )d\tau ]g)=\int \overline {f}(\omega
(s))g(\omega (t))exp[-\int_s^tc(\omega (\tau ),\tau )d\tau ]\; d\mu
_0(\omega )}\eqno (63)$$
$$Q_{s,t} =[\omega ;\: \int_s^t c(\omega (\tau ),\tau )
d\tau <\infty ]$$
The finiteness condition $\int_s^tc(\omega (\tau ),\tau )d\tau
<\infty
$, surely does not hold true, \cite{klaud2}, if $\delta >0$ is
sufficiently small, cf. (60),(61).

Let us mention that some  interesting
mathematical questions were  left aside in the present discussion.
 For example, even in case of conventional
Feynman-Kac kernels, the weakest possible  criterions allowing for
their continuity in spatial variables  are not yet established.
An issue of
the continuity of the kernel in case of general singular potentials,
needs an investigation as well. The phenomenological recipes
 justifying the concrete choice of the kernel are not
 unequivocally established as yet,
 see e.g. our discussion of the Smoluchowski drift
versus Feynman-Kac potential issue in Ref. \cite{blanch}.

{\bf Acknowledgement}: The author receives  a financial support from
the KBN research grant No 2 P302 057 07.

\end{document}